\documentclass[[aps,twocolumn,showpacs,amsmath,amssymb]{revtex4-1}

\usepackage{bm}
\usepackage{epsfig}
\usepackage{graphicx}
\usepackage{booktabs}
\usepackage{multirow}
\usepackage{tabularx}
\usepackage{times}
\usepackage{color}
\usepackage{subfigure}
\usepackage{todonotes}
\usepackage{ulem}
\usepackage[english]{babel}

\definecolor{Red}{rgb}{1,0,0}

\definecolor{Blu}{rgb}{0,0,01}

\definecolor{Green}{rgb}{0,1,0}

\newcommand{\be}{\begin{equation}}
\newcommand{\ee}{\end{equation}}

\newcommand{\ie}{{i.e.},~}
\newcommand{\eg}{{e.g.}~}

\begin{document}

\title{Long-ranged triplet supercurrent in a single in-plane ferromagnet with spin-orbit coupled contacts to superconductors}

\author{Johannes R. Eskilt}

\affiliation{Center for Quantum Spintronics, Department of Physics, Norwegian
University of Science and Technology, NO-7491 Trondheim , Norway}

\author{Morten Amundsen}

\affiliation{Center for Quantum Spintronics, Department of Physics, Norwegian
University of Science and Technology, NO-7491 Trondheim, Norway }

\author{Niladri Banerjee}

\affiliation{Department of Physics, Loughborough University, Loughborough,
LE11 3TU, United Kingdom}

\author{Jacob Linder}

\affiliation{Center for Quantum Spintronics, Department of Physics, Norwegian
University of Science and Technology, NO-7491 Trondheim, Norway}

\begin{abstract}
By converting conventional spin-singlet Cooper pairs to polarized spin-triplet pairs, it is possible to sustain long-ranged spin-polarized supercurrents flowing through strongly polarized ferromagnets. Obtaining such a conversion via spin-orbit interactions, rather than magnetic inhomogeneities, has recently been explored in the literature. A challenging aspect with regard to experimental detection has been that in order for Rashba spin-orbit interactions, present \eg at interfaces due to inversion symmetry breaking, to generate such long-ranged supercurrents, an out-of-plane component of the magnetization is required. This limits the choice of materials and can induce vortices in the superconducting region complicating the interpretation of measurements. Therefore, it would be desirable to identify a way in which Rashba spin-orbit interactions can induce long-ranged supercurrents for purely in-plane rotations of the magnetization. Here, we show that this is possible in a lateral Josephson junction where two superconducting electrodes are placed in contact with a ferromagnetic film via two thin, heavy normal metals. The magnitude of the supercurrent in such a setup becomes tunable by the in-plane magnetization angle when using only a single magnetic layer. These results could provide a new and simpler way to generate controllable spin-polarized supercurrents than previous experiments which utilized complicated magnetically textured Josephson junctions.
\end{abstract}

\date{\today}

\maketitle

\section{Introduction}

\begin{figure}[h]
    \centering
    \includegraphics[width=0.999\columnwidth]{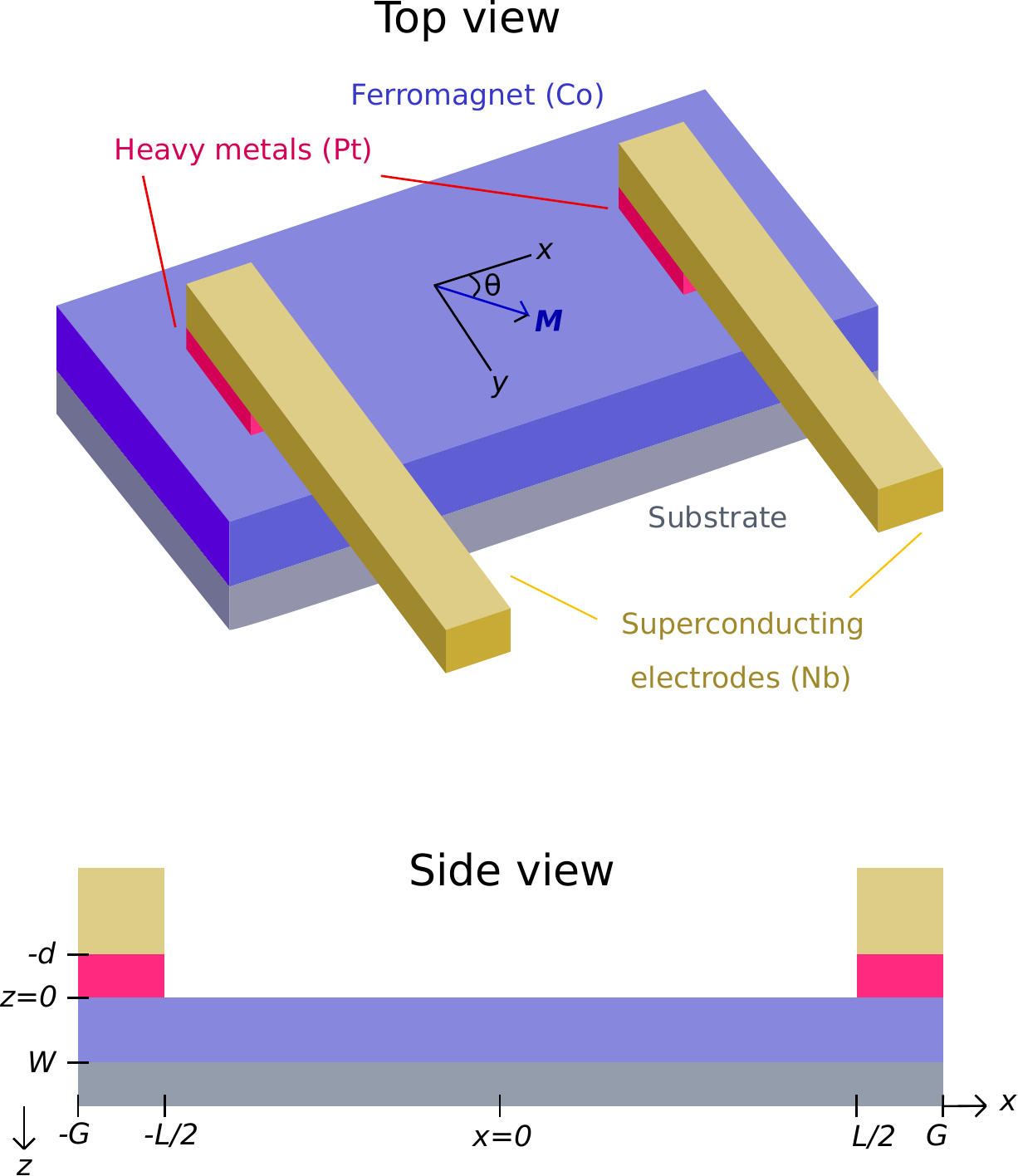}
    \caption{A lateral SFS Josephson junction with Rashba spin-orbit coupling in the heavy metals. The exchange field lies in the plane of the ferromagnet. A supercurrent is sent through the magnetic layer via the superconducting electrodes and is tuned via the in-plane angle $\theta$ of the ferromagnet. Possible choices of materials for the various layers are indicated in the figure. We emphasize that by lateral geometry, we mean a geometry where the superconducting electrodes are attached on top of the sides of the film through which the supercurrent passes, as shown in the figure. This should be viewed in contrast to a \textit{stacked} geometry in which case the structure is built up of successive layers placed on top of each other.}
    \label{fig:model}
\end{figure}

When a conventional superconductor is placed in proximity to a normal state metal, the Cooper pairs will start leaking across the interface from the superconductor and into the metal. These singlet superconducting correlations will, in the metal, start decaying over a length scale of $\xi_N = \sqrt{D/T}$ where $D$ is the diffusion constant of the metal and $T$ is the temperature \cite{likharev1976}. If the metal is a ferromagnet, then the anti-parallel electrons of the singlet Cooper pair will be injected into two different sub-bands (majority and minority) in the ferromagnet, making their Fermi momenta different. This makes the pair decay even faster, namely on a length scale of order $\xi_F = \sqrt{D/M}$ where $M$ is the amplitude of the exchange field. This pair breaking effect can be avoided if the singlet pair can be converted into a triplet pair with a non-zero spin projection along the exchange field. With these so-called long-ranged triplets (LRTs), the pairs will decay slower and be comparable to correlation lengths of normal metal $\xi_N$. Physical quantities like supercurrents will be on the same order, and it is thus of great interest to be able to manipulate and create such LRTs. This topic is currently under intense focus \cite{linder2015, eschrig_rpp_15} because of the potential to develop not only cryogenic spintronics devices, but also radically novel theoretical and experimental aspects of how such pairs can be generated and tuned in a controllable manner.

It is well known theoretically and experimentally that LRT components can be created in ferromagnets where the exchange field has an inhomogenous orientation. Such an exchange field can either be intrinsic to the ferromagnet, or can be fashioned artificially by stacking several layers of homogoneous ferromagnets with misaligned exchange fields \cite{robinson_science_10, khaire_prl_10, bergeret_rmp_05, banerjee_natcom_14,wang_prb_14, leksin_prl_12}. The former severely limits the selection of materials, making the latter more feasible for practical applications. On the other hand, stacks of misaligned ferromagnets present their own challenges, particularly in terms of exerting control over the triplet generation, as manipulating the relative angle between ferromagnetic layers can be difficult.  
This is what makes the discovery that spin-orbit coupling can act as a source of LRTs \cite{bergeret2014} so promising. Indeed, the presence of LRTs in superconductor-ferromagnet structures with only a single homogeneous ferromagnet has been theoretically predicted if heavy metal layers are introduced to the system. This requires, however, that the exchange field in the ferromagnetic layer has both an in-plane and an out-of-plane component~\cite{ali2015}. 
Although such a scenario is possible to obtain experimentally \cite{banerjee_prb_18, satchell_prb_18, satchell_arxiv_19}, it complicates the unambiguous identification of spin-polarized Cooper pairs due to the additional flux injection into the superconductor from the ferromagnet and also severely restricts the choice of materials showing a tailored out-of-plane anisotropy. Recently, further experimental corroboration for the generation of LRTs in superconductor-ferromagnet hybrid structures due to the presence of interfacial spin-orbit coupling has been found in magnetoresistance measurements~\cite{martinez_arxiv_2018}, and in spin pumping experiments~\cite{jeon_nm_2018, jeon_prb_2019, jeon_arxiv_2019}. However, LRTs remain undetected in Josephson junctions. In fact, Refs.~\cite{satchell_prb_18, satchell_arxiv_19} found no clear signature of a long ranged triplet supercurrent in a Josephson weak link with heavy metal layers attached to a ferromagnet with an effective canted magnetization direction, which is in stark contrast to theoretical predictions~\cite{jacobsen_scirep_16}.

It has been proposed \cite{bergeret2014} that lateral geometries may provide less stringent requirements to generate LRTs compared to a stacked geometry, which has the potential to ease their experimental detection in Josephson junctions. 
In particular, it would be desirable to identify a setup where the LRTs can be tuned with a solely in-plane variation of the magnetization, in order to minimize the stray field effect on the superconductor itself. This would be a different result than previous works \cite{jacobsen_scirep_16, nakhmedov_prb_17, satchell_prb_18, satchell_arxiv_19, costa_prb_17,  buzdin_prl_08, arjoranta_prb_16} that have considered how to control the supercurrent via magnetization in Josephson contacts with spin-orbit coupling. A long-ranged supercurrent was predicted in Ref. \cite{bergeret2014}, but without any accompanying study of its dependence on the magnetization direction in the ferromagnetic film.

In this paper, we consider a lateral Josephson junction where two superconducting electrodes are placed in contact with a ferromagnetic layer through a heavy metal (see Fig. \ref{fig:model}). Due to the inversion symmetry breaking and the large atomic number of such metals, Rashba spin-orbit coupling is assumed to be present at these interfaces. As we will show, such a setup will not only host long-ranged triplet Cooper pairs but also give a long-ranged supercurrent only for certain in-plane rotations of the exchange field. Thus, the supercurrent in the ferromagnet is extremely sensitive to this in-plane rotation as long as there is a non-zero spin-orbit coupling present in the heavy metals. We will also show that for some parameters, the in-plane rotation is be able to create $0-\pi$ transitions, which means that for a certain in-plane rotation of the exchange field, the supercurrent is zero. Therefore, such a geometry can work as a transistor for supercurrents by simply rotating the in-plane magnetization. We emphasize that the main novelty and benefit of the present result and setup compared to previous works is that the supercurrent is tuned with a single ferromagnetic layer \textit{and} the magnetization only needs to rotate in the plane of the magnet. Experimental observation of this effect would represent a significant advance with regard to simplifying control over long-ranged spin-polarized supercurrents, which has proved challenging before \cite{banerjee_natcom_14}.

\section{Theory}

In this paper we will use the quasiclassical theory of superconductivity \cite{belzig1999, rammer1986} and consider the dirty limit so that the quasiclassical Green's function $\check{g}$ in the ferromagnet can be described by the Usadel diffusion equation \cite{usadel1970}

\begin{equation}
    i D \Tilde{\nabla} \cdot \left( \check{g} \Tilde{\nabla} \check{g}\right) = [\epsilon \hat{\rho}_3 + \textbf{\textit{M}} \cdot \hat{\bm{\sigma}}, \check{g}]_-,
\end{equation}

where $D$ is the diffusion constant for the ferromagnet, $\epsilon$ is the energy of the quasiparticles, $\hat{\rho}_3 = \text{diag}(1, 1, -1, -1)$,  and $\textbf{\textit{M}}$ is the exchange field. The Pauli matrix vector is $\hat{\bm{\sigma}} = \text{diag}(\underline{\bm{\sigma}}, \underline{\bm{\sigma}}^*)$. The Green's function $\check{g}$ is the $8\times 8$ Green's function in Keldysh space

\begin{equation}
    \check{g} = \begin{bmatrix}
    \hat{g}^R & \hat{g}^K\\
    0 & \hat{g}^A
    \end{bmatrix}.
\end{equation}

Due to the triangular structure of $\check{g}$, the Usadel equation becomes the same for the retaraded Green's function $\hat{g}^R$.

To incorporate spin-orbit coupling into our theory, we have defined \cite{bergeret2014} $\Tilde{\nabla}(\cdot) = \nabla (\cdot) - i [\hat{\bm{A}}, (\cdot)]_-$. Here, $\hat{\bm{A}} = \text{diag}(\underline{\bm{A}}, -\underline{\bm{A}}^*)$, where $\underline{\bm{A}}$ is a $2\times 2$ matrix in spin space which couples to the momentum $\bm{k}$. In effect, the spin-orbit coupling is included as an effective SU(2) gauge-like field, which is possible if it is linear in momentum. We will include both Rashba and Dresselhaus effects in this paper denoted by their respective constants $\alpha$ and $\beta$, both being precisely linear in momentum. However, we emphasize that the main merit of the present setup is that \textit{only} Rashba spin-orbit coupling and an in-plane rotation of the magnetization is required to get a tunable long-ranged supercurrent. The Dresselhaus term is thus simply included to make the results more general. 
Rashba spin-orbit coupling can arise from the lack of inversion symmetry at the interface between two materials. We will later consider two heavy metals where the width in the $z$-direction is small, and thus the Rashba Hamiltonian is of the form

\begin{equation}
    H_R = \frac{\alpha}{m}\left(k_x \underline{\sigma}_y - k_y \underline{\sigma}_x \right),
\end{equation}

where $\bm{k}$ is the momentum of the quasiparticles. The Dresselhaus SOC, on the other hand, can be caused by lack of inversion center in the crystal structure. For two dimensional structures in the $xy$-plane this Hamiltonian becomes

\begin{equation}
    H_D = \frac{\beta}{m}\left(k_y \underline{\sigma}_y - k_x \underline{\sigma}_x \right).
\end{equation}

A term proportional to $\sigma_z$ can appear if the heavy metal has a noncentrosymmetric crystal structure and additionally is not confined along one axis, i.e. not thin. This is a different scenario than the one considered in this work. The two Hamiltonians above can be incorporated into the Usadel equation by using the 2x2 spin-space vector potential $\bm{\underline{A}}$. This gives us

\begin{equation}
    \underline{\bm{A}} = \left(\beta\underline{\sigma}_x -\alpha \underline{\sigma}_y  \right)\bm{e}_x +\left(\alpha\underline{\sigma}_x -\beta \underline{\sigma}_y  \right)\bm{e}_y.
\end{equation}

We will complement the Usadel diffusion equation with Kupriyanov-Lukichev (KL) boundary conditions \cite{kupriyanov1988}

\begin{equation}
    \label{klbc}
    2\zeta L \check{g}\Tilde{\nabla} \check{g} = [\check{g}_l, \check{g}_r]_-,
\end{equation}

where $l$ and $r$ denotes the left and right side of the interface, respectively. $L$ is the length of the respective materials and $\zeta$ is the ratio between the barrier resistance and the bulk resistance. Here, we have also added the gauge covariant derivative $\Tilde{\nabla}$ to include spin-orbit coupling. The interface parameter $\zeta$ depends on microscopic parameters such as the normal-state conductivity and its magnitude determines the magnitude of the proximity-induced superconducting correlations. Our results do not change qualitatively when varying the strength of $\zeta$ and we thus choose to treat it as a phenomenological parameter rather than specifying the exact value of the normal-state conductivities and resistance of the interface regions.

\medskip

To calculate the supercurrent going through the ferromagnetic bridge, we will use quasiclassical expression for the electric current, following the notation of \cite{gomperud2015} and \cite{ali2015}

\begin{equation}
    I_Q = \frac{N_0 D A e}{4} \int^{\infty}_{-\infty} d\epsilon \text{Tr}\left(\hat{\rho}_3 \left(\check{g}\Tilde{\nabla} \check{g} \right)^K \right).
\end{equation}

Here, $A$ is the cross section, $N_0$ is the density of states at Fermi level, $e$ is the electric charge. The superscript $K$ denotes the Keldysh component of the $8\times 8$ matrix. The system in consideration will be in equilibrium, and thus we can use the relation $\hat{g}^K = \tanh(\beta \epsilon/2)\left(\hat{g}^R-\hat{g}^A \right)$ where $\beta$ in this context is the inverse temperature $1/k_BT$ and should not be confused with the Dresselhaus constant. The expression for the charge supercurrent then takes the form

\begin{equation}
    I_Q = I_0 \int^{\infty}_{-\infty} d\epsilon \tanh(\beta \epsilon/2)\text{Tr}\left(\hat{\rho}_3 \left(\hat{g}^R\Tilde{\nabla} \hat{g}^R - \hat{g}^A \Tilde{\nabla}\hat{g}^A \right)\right),
\end{equation}

where $I_0=\frac{N_0 D A e}{4}$. We can find $\hat{g}^R$ with the Usadel equation, and with the relation $\hat{g}^A = - \hat{\rho}_3 \left(\hat{g}^R \right)^{\dagger}\hat{\rho}_3$, we have everything we need to find the supercurrent. Later, we will compare our result with the supercurrent through a ferromagnetic film when no interfacial spin-orbit coupling is present. In this case, the derivatives become normal derivatives \ie $\Tilde{\nabla} \rightarrow \nabla$. It can easily be shown that this current is conserved in regions that are governed by the Usadel equation, both with and without spin-orbit coupling \ie $\nabla \cdot I_Q = 0$ \cite{ali2015}. Thus the supercurrent in ferromagnetic region in Fig. \ref{fig:model} will be conserved. 

Our problem is inherently two-dimensional, but we will make it effectively one-dimensional by assuming that the total width of the heavy metals and ferromagnetic film $W+d$ is much smaller than length scale over which the Green's function varies. Thus, we can assume the Green's function stays roughly constant along the $z$-axis, and by averaging the condensate function along the $z$-axis we can apply the KL boundary condition at the superconductor/heavy-metal interfaces. This effectively gives the differential equations a source of singlet Cooper pairs in the two regions $-G<x<-L/2$ and $L/2<x<G$.

We need to solve three sets of differential equations: the condensate functions in the two superconducting nodes and in the ferromagnetic bridge. In order to get an exact solution we need an appropriate set of boundary conditions. At the two vacuum interfaces, $x=-G$ and $x=G$, we use the KL boundary conditions. At $x=-L/2$ and $x=L/2$ we require that the condensate functions are continuous and we also use the KL boundary conditions here. Since the two condensate functions must be continuous, the right side of the KL boundary condition in equation \eqref{klbc} will be zero. And thus, we effectively get the condition

\begin{equation}
    \label{contkl}
    \check{g}_i\Tilde{\nabla}\check{g}_i = \check{g}_j\Tilde{\nabla}\check{g}_j,
\end{equation}

where $(i, j)$ are two regions in contact. This also ensures supercurrent conservation.

\section{Results and Discussion}

\subsection{Analytical results: general considerations}

Before resorting to a numerical analysis, we can draw several conclusions by making use of the weak proximity effect approximation. The assumption is that in any non-superconducting materials, the Cooper pair correlations will be weak, and thus the retarded Green's function only slightly deviates from its normal-state value:

\begin{equation}
    \label{weakg}
    \hat{g} =
    \begin{bmatrix}
        \underline{1}& \underline{f}\\
        -\underline{\Tilde{f}} & -\underline{1}
    \end{bmatrix},
\end{equation}

where the tilde conjugation $\Tilde{(\cdot)}$ changes the sign of the energy and complex conjugates. We insert this $4\times4$ Green's function matrix to the Usadel equation, and by looking exclusively at the top-right $2\times2$ element, we will get an equation that is completely independent of $\underline{\Tilde{f}}$. Thus, we only need to solve for the four elements in $\underline{f}$ and to get $\underline{\Tilde{f}}$ we perform the tilde-conjugation, i.e. change sign of the energy and complex conjugate.

By applying the weak proximity approximation to the Usadel equation, we can linearize it in the anomalous Green's function $\underline{f}$ to obtain

\begin{align}
    \label{linearusadel}
    \nabla ^2 \underline{f} &- 2i \left[\underline{\bm{A}}, \nabla \underline{f}\right]^*_+ - \left[\underline{\bm{A}}, \left[\underline{\bm{A}}, \underline{f}\right]^*_+\right]^*_+ \notag\\
		&+ \frac{2\epsilon i}{D}\underline{f} + \frac{i}{D}\bm{M}\cdot\left[\underline{\bm{\sigma}},\underline{f}\right]^*_-=0.
\end{align}

where we have used the notation $[A, B]^*_+ = AB+BA^*$. We now proceed to show that the KL boundary condition provide an effective source of singlet pairs in our linearized Usadel equation. We will make the standard simplifying assumption that the inverse proximity effect can be neglected and the Green's function in the superconductor is the BCS bulk solution given as

\begin{equation}
    \label{BCSG}
    \hat{g} = \begin{bmatrix}
    \cosh(\theta) & i\sigma^y\sinh(\theta)e^{i \phi} \\
    -i\sigma^y\sinh(\theta)e^{-i\phi} & -\cosh(\theta)
    \end{bmatrix},
\end{equation}

where $\theta = \theta(\epsilon) = \text{atanh}(\Delta/\epsilon)$. We then average over the $z$-direction, which causes the KL boundary condition to act as a source of singlet state pairs in the linear Usadel equation.  Inserting the weak proximity Green's function for the ferromagnetic region and the BCS bulk Green's function, we get

\begin{equation}
    \frac{\partial \underline{f}}{\partial z}-i \left[\underline{A}_z,\underline{f} \right]^*_+|_{S/F} = \frac{ \cosh(\theta)}{\zeta L}\underline{f}-\frac{\sinh(\theta)}{\zeta L}e^{i\phi} i\underline{\sigma}^y.
\end{equation}

As we already have seen, $\underline{A}_z=0$. Since we are assuming that the elements of $\underline{f}$ are much smaller in magnitude than unity, the first term on the right-hand side can be neglected. We will now use this boundary condition by first expanding the Laplace operator $\nabla^2 \underline{f} = \frac{\partial^2 \underline{f}}{\partial x^2} + \frac{\partial^2 \underline{f}}{\partial z^2}$, integrate over the $z$-direction and use the KL boundary conditions,

\begin{equation}
    \int_{-d}^W \frac{\partial^2 \underline{f}}{\partial z^2} \text{d}z = \frac{\partial \underline{f}}{\partial z}|_{z=W} - \frac{\partial \underline{f}}{\partial z}|_{z=-d} = \frac{\sinh(\theta)}{\zeta (W+d)}e^{i\phi} i\underline{\sigma}^y.
\end{equation}

Here, we used that the length normal to the interface is simply $W+d$. By now averaging over all components in the linear Usadel equation, we get

\begin{align}
    \label{linearusadellateral}
    \nonumber
    &\frac{\partial^2}{\partial x^2} \underline{f} - \frac{2id}{W+d} \left[\underline{\bm{A}}, \frac{\partial}{\partial x} \underline{f}\right]^*_+ - \frac{d}{W+d}\left[\underline{\bm{A}}, \left[\underline{\bm{A}}, \underline{f}\right]^*_+\right]^*_+
    \\
    &+ \frac{\sinh(\theta)}{\zeta (W+d)^2}e^{i\phi} i\underline{\sigma}^y
    + \frac{2\epsilon i}{D}\underline{f} + \frac{i}{D}\bm{M}\cdot\left[\underline{\bm{\sigma}},\underline{f}\right]^*_-=0.
\end{align}

This equation has to be solved in three regions, the two superconducting nodes \ie $L/2<x<G$ and $-G<x<-L/2$, and in the ferromagnetic bridge \ie $-L/2<x<L/2$. In the ferromagnetic bridge, we have no spin-orbit coupling and we can simply set $\bm{\underline{A}}=0$ in this region. In the superconducting nodes, the effective magnetization $\bm{M}$ will be smaller than in the ferromagnetic film since there is no exchange field present in the heavy metals, so the effective exchange field is thus $\bm{M} \rightarrow \frac{W}{d+W}\bm{M}$, assuming similar normal-state conductivities of the spin-orbit coupled and ferromagnetic layers. We also allow for different macroscopic phases for the nodes such that the phase difference is $\Delta \phi = \phi_R - \phi_L$. Note that we have performed the standard approximation of neglecting the inverse proximity effect in the superconductors which is formally valid under the assumption that there exists a strong mismatch between the normal-state conductivity of the superconducting material compared to the other layers.

\medskip

Before solving equations, we have to know our boundary conditions. This two-dimensional problem is solved by making the problem effectively one-dimensional, and thus we apply the KL boundary conditions at the vacuum interfaces $x=-G$ and $x=G$ which effectively sets the current moving in the $x$-direction to zero at these edges. At the two interfaces between the three regions, $x=-L/2$ and $x=L/2$, we require that the Green's functions are continuous and the KL boundary condition is satisfied. As mentioned, since the Green's functions are continuous we get equation \eqref{contkl}. In the weak proximity limit, this gives us

\begin{align}
    \label{bc1}
    \partial_x \underline{f}(-L/2^+) &= \partial_x \underline{f}(-L/2^-) -\frac{d}{W+d}i\left[\underline{A}_x, \underline{f}(-L/2^-) \right]^*_+\\
    \label{bc2}
    \partial_x \underline{f}(L/2^-) &= \partial_x \underline{f}(L/2^+) -\frac{d}{W+d}i\left[\underline{A}_x, \underline{f}(L/2^+) \right]^*_+.
\end{align}

For the anomalous Green's function $\underline{f}$, we will make use of the so-called $d$-vector formalism \cite{leggett1975} where all triplet correlations are compactly expressed through a vector $\bm{d}$. The total superconducting anomalous Green's function matrix may then be written as:

\begin{equation}
    \underline{f} = \left(f_s + \bm{d}\cdot \underline{\bm{\sigma}} \right)i\sigma^y = \begin{bmatrix}
        id_y - d_x & d_z + f_s \\
        d_z-f_s & id_y + d_x
    \end{bmatrix}.
\end{equation}

The $d$-vector representation has the advantage of clearly separating the long-ranged and short-ranged triplet component of $\underline{f}$ \cite{ali2015}. The long-ranged component will be component that is perpendicular to the exchange field $d_{LRC} = |\bm{d} \times \bm{M}|$ while the short-ranged component is parallel to the exchange field $d_{SRC} = \bm{d} \cdot \bm{M}$. We can now enter our $d$-vector into equation \eqref{linearusadellateral}. The set of Pauli matrices with the addition of the identity matrices form a basis for a general $2\times 2$ matrix. Therefore, by using the identity $\sigma^a \sigma^b = \delta_{ab}\underline{I} + i\epsilon_{abc}\sigma^c$, we get four equations for each of the four matrices:

\begin{equation}
    \label{Mfs}
    \frac{\partial^2 f_s}{\partial x^2} + \frac{\sinh(\theta)}{\zeta (W+d)^2}e^{i\phi} +\frac{2\epsilon i}{D}f_s+\frac{2i}{D}\left(M_x d_x+ M_y d_y\right) = 0,
\end{equation}

\begin{align}
    \nonumber
    \frac{\partial^2 d_x}{\partial x^2} + \frac{d}{W+d}\bigg(-4\alpha \frac{\partial d_z}{\partial x} &-
    4(\alpha^2+\beta^2)d_x-8\alpha\beta d_y\bigg)\\
    \label{Mdx}
    &+\frac{2\epsilon i}{D}d_x+\frac{2iM_x}{D}f_s=0,
\end{align}

\begin{align}
    \nonumber
    \frac{\partial^2 d_y}{\partial x^2} + \frac{d}{W+d}\bigg(-4\beta \frac{\partial d_z}{\partial x} &- 4(\alpha^2+\beta^2)d_y-8\alpha\beta d_x\bigg)\\
    \label{Mdy}
    &+\frac{2\epsilon i}{D}d_y+\frac{2iM_y}{D}f_s=0,
\end{align}

\begin{align}
    \nonumber
    \frac{\partial^2 d_z}{\partial x^2} + \frac{d}{W+d} \bigg(4\alpha \frac{\partial d_x}{\partial x} &+  4\beta \frac{\partial d_y}{\partial x}- 8(\alpha^2+\beta^2)d_z\bigg)\\
    \label{Mdz}
    &+\frac{2\epsilon i}{D}d_z=0.
\end{align}

We can immediately draw several conclusions before attempting to solve the differential equations. First of all, the transformation $d_x \leftrightarrow d_y, \alpha \leftrightarrow \beta, M_x \leftrightarrow M_y$ leaves the equations invariant. We will mostly look at the case where we only have Rashba spin-orbit coupling present since this case is experimentally more feasible, but due to this invariance, our conclusions of the supercurrent and triplets will also be invariant to this transformation.

We continue by looking at the case $\beta = M_y = 0$, and $M_x \neq 0$. This decouples the third equation from the rest of the equations, and thus there is no way for the singlet state $f_s$ to be transformed into a triplet $d_y$ state. In a spatially homogeneous system, the long ranged triplet state ${d_{LRC} = |\bm{d}\times\bm{M}|\propto d_z}$ decouples as well, and hence only the short ranged triplets $d_x$ emerge. If, on the other hand, there is an uneven distribution of the triplet correlations, this may lead to a precession of the triplet Cooper pairs due to the Rashba spin--orbit coupling. In particular, $\frac{\partial d_x}{\partial x}\neq 0$ causes a precession about the $y$ axis, and the generation of $d_z$ triplets. This is precisely the case for the lateral geometry of Fig.~\ref{fig:model}; the superconducting correlations are largest directly beneath the superconducting electrodes, and reduces in strength as one moves along the $x$ axis, towards $x = 0$, producing the necessary gradient.  

While increasing $\alpha$ increases the production of long ranged triplets, a larger $\alpha$ also has a detrimental effect on all triplet Cooper pairs due to the Dyakonov-Perel-like spin relaxation~\cite{bergeret2014}. The manifestation of this effect is the appearance of an imaginary term in the quasiparticle energy, which for the long ranged triplets takes the form,

\begin{equation}
    \epsilon_{LRT} = \epsilon - i\frac{4dD}{W+d}\alpha^2.
\end{equation}

Imaginary contributions to the energy are normally associated with pair-breaking processes, and therefore, these LRT components will decay faster if the Rashba coefficient is large. On the other hand, if the Rashba coefficient is zero, then there will be no LRTs at all. We therefore expect to find a maximum value for the triplets and supercurrent for a certain intermediate value of $\alpha$. We will later show numerically that this reasoning is correct, resulting in a non-monotonic behavior of the supercurrent as a function of $\alpha$, and that an in-plane rotation of the exchange field will drastically change the magnitude of the supercurrent.


If we instead set $\beta=M_x=0$, and $M_y\neq 0$, we decouple the second and fourth differential equations from the other two, and thus $d_x=d_z=0$. The Rashba coupling has in this case a very small impact on the system and will only impact singlet pairs and the short range triplets (SRTs) with no LRTs present. Thus, in the case of Rashba coupling, an in-plane rotation of the exchange field from $M_x\bm{e}_x$ to $M_y\bm{e}_y$ will make all LRTs vanish and only SRTs will remain.


\subsection{Analytical results: in-plane magnetization}

We will now show explicitly that we get a long-ranged triplet pair correlation with spin-orbit coupling which in turn gives a long-ranged charge-supercurrent. We will only be looking at a pure Rashba spin-orbit coupling and set $\beta=0$. We will also place the magnetic field in the $x$-direction and thus $M_y=0$.

We assume now that the distance $L$ between the two superconducting electrodes is so large that the solution for the anomalous Green's function in the ferromagnetic bridge will consequently be the superposition of the Green's function in two systems with only one effective superconducting node. In this way, we only need to solve the anomalous Green's function in a lateral geometry with one effective superconducting node with spin-orbit coupling present.
Thus, we start by finding the solution for an effective bilayer in which a superconductor with spin--orbit coupling is located in the region $x\leq 0$, and a ferromagnet at $x\geq0$. Far into the semi-infinite regions the solutions will converge to zero, and we only take into account the boundary conditions at $x=0$ in Eqs. \eqref{bc1} and \eqref{bc2} with the addition of continuity of the anomalous Green's functions.
 We will also assume that the Rashba coupling is weak, $\alpha^2 \ll |M|/D$, so that we can remove any second order term in $\alpha$ in the differential equation. The general solution of the differential equations then becomes

\begin{align}
    \nonumber
    f_s &= -\frac{2\alpha k}{K_p^2-k^2}C_4e^{kx}+C_5 e^{K_px}+C_6e^{K_mx} \\
    &+ \frac{k^2}{K_p^2\left(2k^2-K_p^2 \right)}he^{i\phi_1}\\
    d_x &= C_5e^{K_px}-C_6e^{K_mx}-\frac{K_p^2-k^2}{K_p^2\left(2k^2-K_p^2 \right)}h e^{i\phi_1} \\
    d_y &= 0\\
    d_z &= C_4 e^{kx}-\frac{2\alpha K_p}{K_p^2-k^2}C_5e^{K_px}-\frac{2\alpha K_m}{K_p^2-k^2}C_6e^{K_mx}
\end{align}

 when $x < 0$. Here, $k=\sqrt{-2i\epsilon/D}$, $K_{p(m)}=\sqrt{-2i(\epsilon+(-)M_x)/D}$ and $h=\sinh(\theta)/{\zeta (W+d)^2}$ and in the ferromagnetic bridge when $x>0$ the solution is

\begin{align}
    f_s &= -C_1e^{-K_m x} + C_2 e^{-K_p x}\label{eq:fs0} \\
    d_x &= C_1e^{-K_m x} + C_2 e^{-K_p x}\label{eq:dx0}\\
    d_z &= C_3e^{-kx}.\label{eq:dz0}
\end{align}

As expected, only $d_z$ has any long-ranged triplet components in the purely ferromagnetic region, and thus we are mostly interested in finding $C_3$. Applying the boundary conditions at $x=0$, we get to the first order in $\alpha$,

\begin{align}
    \nonumber
    C_3 &= -\frac{3K_p^4-kK_p^3 + (k K_m -6k^2)K_p^2+2k^3K_p+k^4}{\left(4kK_p^6-12k^3K_p^4+8k^5K_p^2\right)}\\
    \label{wholec3}
    &\times\frac{d}{W+d} \alpha he^{i\phi_L},
\end{align}

which clearly shows that we only get a long-ranged triplet component if we have Rashba spin-orbit coupling present. Letting $|M_x| \gg \epsilon$, we get $|K_{(m/p)}| \gg |k|$ and

\begin{equation}
    \label{approxC3}
    C_3 = -\frac{3d\alpha h e^{i\phi_L}}{4(W+d)kK_p^2}.
\end{equation}

We now place a second superconducting electrode at $x=L/2$ and push the first electrode back to $x=-L/2$. 
We solve the differential equations for the second node and assume that total condensate function $\underline{f}$ is a superposition of the two solutions and that the superconducting nodes are so far apart that the overlap between the two solutions is small. The complete solution for the long-ranged component is thus

\begin{align}
    d_z &= C_3^-e^{-k(x+L/2)}+C_3^+e^{k(x-L/2)}.
\end{align}

Here, $C_3^-$ is the coefficient for the left superconducting node and $C_3^+$ for the other node. $C_3^-$ is given in Eq. \eqref{wholec3}, while $C_3^+$ is found by making the replacements $k\to-k$, $K_{(p/m)}\to-K_{(p/m)}$, and $\phi_L\to\phi_R$ 
Entering this LRT component into the formula for the supercurrent, we get

\begin{align}
    I_Q &= 4N_0 D e\int^{\infty}_{0} d\epsilon \tanh(\beta \epsilon/2)\\
    &\times \Re\bigg(k\left(C_3^+ \Tilde{C}_3^- -C_3^-\Tilde{C}_3^+\right)e^{-kL}\bigg).
\end{align}

Here the tilde conjugation is as mentioned just doing the transformation $\epsilon \rightarrow -\epsilon$ and $i \rightarrow -i$. Using the approximated $C_3$ in Eq. \eqref{approxC3}, the long-ranged supercurrent becomes

\begin{align}
    \nonumber
    I_Q &= 8N_0 D e \sin(\Delta \phi) \int^{\infty}_{0} d\epsilon \tanh(\beta \epsilon/2) \left(\frac{3d\alpha}{4(W+d)}\right)^2\\
    &\times \Re\bigg( -i \frac{h \Tilde{h}}{kK_m^2K_p^2}e^{-kL}\bigg),
\end{align}

where $\Delta \phi = \phi_R-\phi_L$. Therefore, this long-ranged triplet component also gives a long-ranged supercurrent that is proportional to $\alpha^2$ for small $\alpha$. In this expression for the supercurrent, we have used the simplified $C_3$ solution which amounts to the approximation that the main contribution to the integral for the supercurrent comes from the region $\varepsilon \ll |M_x|$. Numerically, we have confirmed that the main contribution indeed comes from the region near $\varepsilon=\Delta$. Alternatively, and more accurately, we could simply use the whole solution for $C_3$ in equation \eqref{wholec3} which results in a much longer expression for $I_Q$. The point is nevertheless that we get a long-ranged supercurrent when $\alpha \neq 0$. As previously argued, if we rotate the exchange field from a pure $x$-direction to lie along the $y$-axis, the long-ranged component will become zero. An in-plane rotation of the exchange field from $\bm{M} = M \bm{e}_x$ to $\bm{M} = M \bm{e}_y$ with Rashba coupling should therefore result in a significant drop in the magnitude of the supercurrent.

As mentioned above, the system is invariant under the transformation $d_x \leftrightarrow d_y, \alpha \leftrightarrow \beta, M_x \leftrightarrow M_y$ and hence we get the same expression for the long-ranged supercurrent with $\beta$ instead of $\alpha$ if we set $\alpha=M_x=0$ and keep $\beta$ and $M_y$ non-zero. This means that pure Dresselhaus spin-orbit coupling would also be sufficient to get a long-ranged supercurrent. 

\subsection{Numerical results}

The weak proximity approximation is only valid if the magnitude of the elements of $\underline{f}$ are much smaller than unity which limits the choice of parameter values that can be explored. We will solve the full proximity effect Usadel equation numerically in this section, which is free from this restriction.

We will solve the problem by using the Riccati parameterization with spin-orbit coupling derived in Ref. \cite{ali2015},

\begin{align}
    \nonumber
    D\left(\nabla^2 \gamma + 2 \left(\nabla \gamma \right)\Tilde{N}\Tilde{\gamma}\left(\nabla \gamma \right) \right) = -2i\epsilon \gamma -i\bm{M}\cdot \left(\bm{\sigma}\gamma -\gamma\bm{\sigma}^* \right)\\
    \nonumber
    +D\left(\underline{\bm{A}}^2\gamma - \gamma \left(\underline{\bm{A}}^*\right)^2+2\left(\underline{\bm{A}}\gamma + \gamma \underline{\bm{A}}^* \right)\Tilde{N}\left(\underline{\bm{A}}^* + \Tilde{\gamma}\underline{\bm{A}}\gamma \right) \right)\\
    +2iD\left(\left(\nabla \gamma \right)\Tilde{N}\left(\underline{\bm{A}}^*+\Tilde{\gamma}\underline{\bm{A}}\gamma\right) + \left(
    \underline{\bm{A}}+\gamma \underline{\bm{A}}^*
    \Tilde{\gamma} \right)N\left(\nabla \gamma \right)
    \right).
	\label{eq:fullusadel}
\end{align}

The corresponding equation for $\Tilde{\gamma}$ can be found by tilde conjugating the equation above. Here, the Green's functions are given as $\underline{g} = N(1+\gamma \Tilde{\gamma})$ and $\underline{f} = 2N\gamma$. And $N = (1-\gamma \Tilde{\gamma})^{-1}$, and thus we need to solve for $\gamma$ and $\Tilde{\gamma}$. We will still be approximating the system to be one dimensional with the KL boundary conditions in the two nodes working as two sources of singlet states. The KL boundary conditions are

\begin{align}
    \frac{\partial}{\partial z} \gamma = \frac{1}{L \zeta}\left(1 - \gamma \Tilde{\gamma}_S\right)N_S\left(\gamma - \gamma_S\right) + i \underline{A}_z\gamma + i \gamma\underline{A}_z^*
\end{align}

where $\zeta$ is the ratio between the barrier resistance and the bulk resistance of the heavy metal, and $L$ is the width of the normal metal and ferromagnetic layer which is $L=W+d$. $\gamma_S$ and $N_S$ are the Riccati parameters for the BCS bulk superconductor. Since the width $W$ of the heavy metal and the ferromagnetic film is small, we will neglect the inverse proximity effect and use the bulk BCS Green's functions in the superconductors. We will as in the last section use this boundary condition between the heavy metal and the superconductor as an effective source of singlet state pairs. Since the normal vector of the interface points in the $z$-direction, we get $\underline{A}_z=0$. The $z$-component of $\nabla^2 \underline{\gamma}$ will be non-zero when averaged over the $z$-direction, and the effective Usadel equation becomes:

\begin{align}
    \nonumber
    &D\bigg[\frac{\partial^2}{\partial x^2} \gamma + \frac{1}{(W+d) \zeta}\left(1 - \gamma \Tilde{\gamma}_S\right)N_S\left(\gamma - \gamma_S\right)\\
    \nonumber
    &+2 \left(\frac{\partial}{\partial x} \gamma \right)\Tilde{N}\Tilde{\gamma}\left(\frac{\partial}{\partial x} \gamma \right) \bigg]\\
    \nonumber
    &= -2i\epsilon \gamma -i\bm{M}\cdot \left(\bm{\sigma}\gamma -\gamma\bm{\sigma}^* \right)\\
    \nonumber
    &+D\frac{d}{W+d}\bigg[\underline{\bm{A}}^2\gamma - \gamma \left(\underline{\bm{A}}^*\right)^2\\
    \nonumber
    &+2\left(\underline{\bm{A}}\gamma + \gamma \underline{\bm{A}}^* \right)\Tilde{N}\left(\underline{\bm{A}}^* + \Tilde{\gamma}\underline{\bm{A}}\gamma \right) \bigg]\\
    \nonumber
    &+2iD\frac{d}{W+d}\bigg[\left(\frac{\partial}{\partial x} \gamma \right)\Tilde{N}_F\left(\underline{\bm{A}}^*+\Tilde{\gamma}\underline{\bm{A}}\gamma\right)\\
    &+ \left(
    \underline{\bm{A}}+\gamma \underline{\bm{A}}^*
    \Tilde{\gamma} \right)N\left(\frac{\partial}{\partial x} \gamma \right)\bigg].
\end{align}

The corresponding equation for $\Tilde{\gamma}$ can be found by tilde conjugation the equation above. By using the bulk BCS Green's function, we can easily calculate $N_S$ and $\gamma_S$.


We consider the system depicted in Fig. \ref{fig:model}. The diffusive limit coherence length of the superconductor is $\xi_S = \sqrt{D/\Delta}$, where $\Delta$ is the superconducting gap energy. We will use the lengths $W/\xi_S=d/\xi_S=0.08$ and $L/\xi_S=1$. We will also let the length of the spin-orbit coupled region be $0.2\xi_S$, which gives us $G/L = 0.7$. The interface transparency will be $\zeta = 5$, and the exchange field is placed in the $xy$-plane $\bm{M} = M (\cos(\theta), \sin(\theta), 0)$. We normalize $\epsilon$ and $M$ to the gap energy $\Delta$. We choose a strong ferromagnet $M_{F} = 50\Delta$ and with $W=d$, the effective exchange field will be $M = 25 \Delta$ in the two superconducting electrodes and $M=50\Delta$ in the middle region. The value of the exchange field is reasonable considering an ultra-thin strong ferromagnet like cobalt in contact with a heavy metal like platinum \cite{banerjee_prb_18}. The macroscopic phase difference has been set to $\Delta \phi = \phi_R - \phi_L = \pi/2$, while the temperature is $T = 0.5 T_C$, and in addition, we will now only assume a pure Rashba coupling which we will normalize to the length of the ferromagnetic bridge $L$ such that $\alpha L$ will be a dimensionless quantity. The spin-orbit coupling term is then

\begin{equation}
    \underline{\bm{A}} = -\alpha\underline{\sigma}_y\bm{e}_x + \alpha \underline{\sigma}_x\bm{e}_y.
\end{equation}

\begin{figure}[h]
\includegraphics[width=\columnwidth]{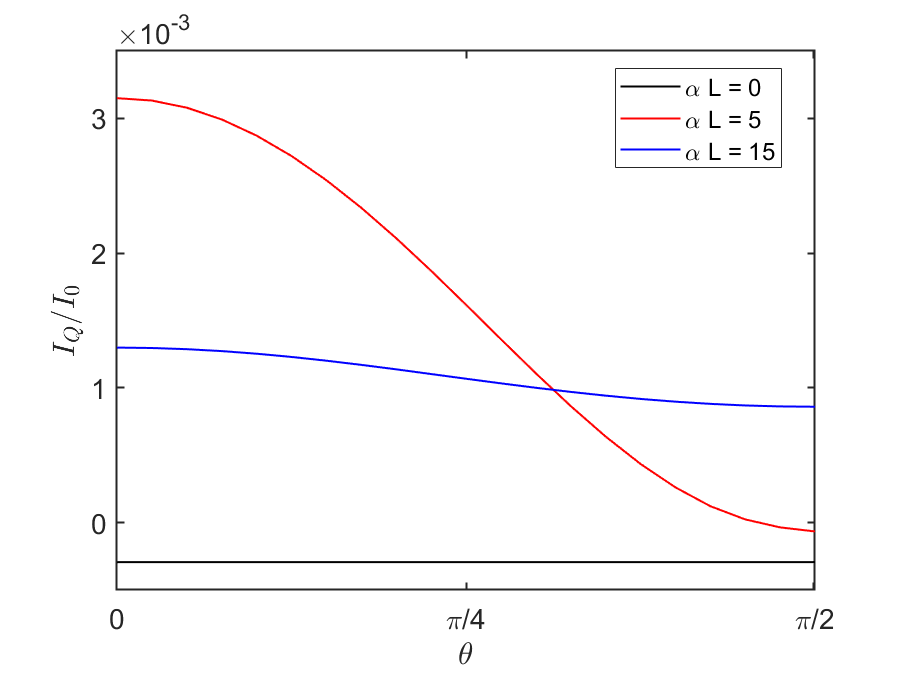}
\caption{The supercurrent is plotted as a function of exchange field $\theta$. When $\theta=0$ the exchange field points along the $x$-direction, while $\theta=\pi/2$ corresponds to the exchange field pointing in the $y$-direction.}
\label{fig:theta}
\end{figure}

\begin{figure}[h]
\includegraphics[width=\linewidth]{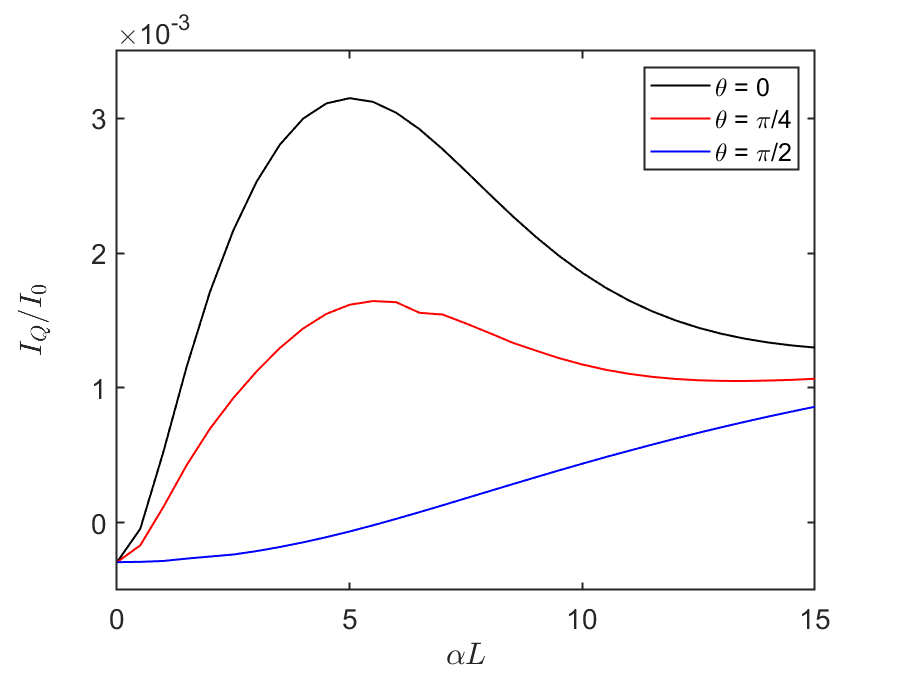}
\caption{The supercurrent is plotted as a function of Rashba coupling $\alpha$ in the heavy metals.}
\label{fig:rashba}
\end{figure}

The supercurrent is plotted as a function of the exchange field angle $\theta$ and Rashba coupling in Fig. \ref{fig:theta} and \ref{fig:rashba}, respectively, where $I_0=N_0 D A e$. With Rashba coupling, we clearly see an enhanced supercurrent when the exchange field points in the $x$-direction ($\theta=0$). There also seems to be a certain magnitude of the Rashba constant where the supercurrent is peaked when $\theta = 0$, namely at $\alpha L \approx 5$. Interestingly, we also see from Fig. \ref{fig:rashba} that we are able to create $0-\pi$ transitions when the strength of the Rashba coupling is $\alpha L \leq 6$. Thus, there exists an angle close to $\theta = \pi/2$ where the current is zero as long as $\alpha L < 6$. It also seems that the supercurrent becomes independent of $\theta$ when $\alpha L \rightarrow \infty$. This is, as we explained in the weak proximity limit, because the energy of the LRTs get an imaginary part which destroy the coherence of these components.

\begin{figure}[h]
\includegraphics[width=\columnwidth]{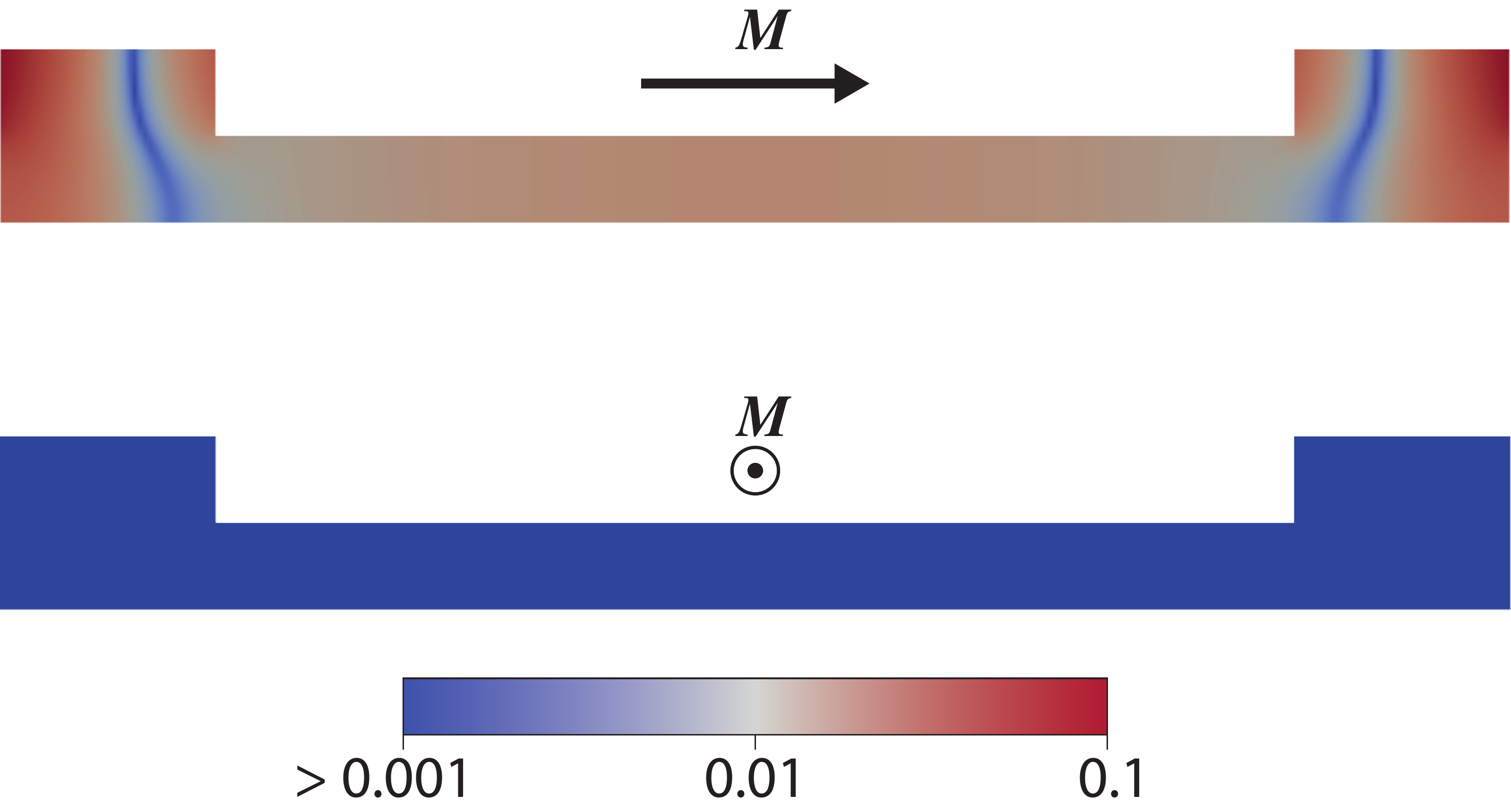}
\caption{The long ranged triplet pair correlation $\Phi_z$ plotted for an exchange field pointing in the $x$ direction (top) and the $y$ direction (bottom). Only the former yields a nonzero $\Phi_z$.}
\label{fig:PC}  
\end{figure}

To more properly understand the behavior of the supercurrent in the system, we compute the triplet pair correlation, which is defined as~\cite{lofwander2005,montiel2018}
\begin{equation}
\bm{\Phi} = \int^\infty_0 d\varepsilon\;\bm{d}\tanh\frac{\beta\varepsilon}{2}.
\label{eq:PC}
\end{equation}
The pair correlation for the long ranged triplet component, $\Phi_z$, for the exact two dimensional geometry considered is computed using a Galerkin finite element method~\cite{amundsen_scirep_2016}, and is shown in Fig.~\ref{fig:PC}. We note in passing that this finite element method divides the system into elements defined by nodes. Some of these nodes are corner nodes, but the actual solution of the partial differential equation within each element requires functional evaluations at points which differ from the nodal points. As a result, one does not have to deal with surface discontinuities at corners. Transparent boundary conditions have been assumed between the heavy metal layers and the ferromagnet. It is seen that when the exchange field is pointing in the $x$ direction, $d_z$ triplets accumulate along the vacuum edges of the heavy metal layers. The reason for this is that the vacuum edges constrain the Cooper pair diffusion in the $x$ direction, giving a nonzero gradient $\frac{\partial d_x}{\partial x}$ in the density of these triplet pairs. Due to the spin--orbit coupling, such a gradient acts as a source for $d_z$ triplets due to spin precession. The corresponding pair correlation leaks into the ferromagnet, and being long ranged with respect to the exchange field, it permeates the entire ferromagnetic bridge, thus acting as a mediator for the supercurrent. In contrast, no $d_z$ triplets are found when the exchange field points in the $y$ direction, which is consistent with the reduction in the magnitude of the supercurrent seen in Fig.~\ref{fig:theta}. It is clear that a finite thickness $d$ of the heavy metal layers is essential for the generation of $d_z$ triplets. This means that models which approximate the spin--orbit coupling as solely an interface effect, e.g., as discussed in Ref.~\cite{amundsen2019}, will fail to capture the correct $\theta$ dependence.

\begin{figure}
\includegraphics[width=\linewidth]{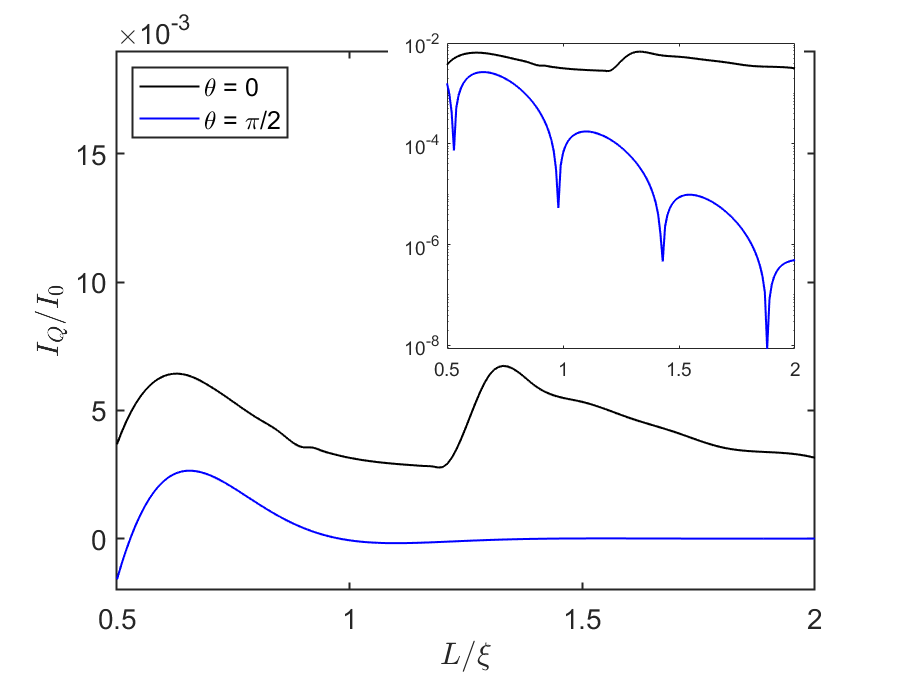}
\caption{The supercurrent plotted as a function of the length of the ferromagnetic bridge $L$. The inset is a log-plot of the absolute value of the current and shows how vastly  different the exponential decay is for the two in-plane directions of the exchange field are. The sharp dips in the log graph shows where the short ranged current switches sign.}
\label{fig:L}
\end{figure}

The key observation is that there should be a change in the critical current of the system with an in-plane magnetization rotation, an effect which is absent in systems with spin-singlet supercurrents. The supercurrent is also plotted as a function of the length of the ferromagnetic region in Fig. \ref{fig:L} where we have set $\alpha \xi = 5$. This choice corresponds to the maximum supercurrent in Fig. \ref{fig:rashba} when $L/\xi = 1$. We see that the supercurrent in the case of a pure $x$-directed exchange field $(\theta=0)$ decays much slower than in the case where the exchange field points along the $y$-axis $(\theta=\pi/2)$. This is precisely due to the fact that the supercurrent is now carried by long-ranged triplet Cooper pairs. Note that the supercurrent rapidly changes sign when $\theta=\pi/2$ due to 0-$\pi$ oscillations. In contrast, for $\theta=0$ there is no $0-\pi$ transitions in the interval $0.5 < L/\xi < 2$. This allows for an interesting observation, namely that there exists several possible intervals of $L/\xi$ where a 90 degree in-plane rotation of the magnetization essentially turns the supercurrent on and off.

\section{Concluding remarks}

We have shown that a lateral Josephson junction with spin-orbit coupled contacts to a ferromagnetic film that is magnetized in-plane is able to carry a long-ranged triplet supercurrent. This supercurrent is highly sensitive to the in-plane rotation of the magnetic field, and our system thus effectively acts as a magnetic transistor for the supercurrent. The main merit of our result is that the long-ranged triplet supercurrent is tuned with a single ferromagnetic layer without any requirement for an out-of-plane magnetization. We believe this could provide a way to realize tunable triplet supercurrents via Rashba spin-orbit coupling in a considerably simpler way than previous proposals.

\acknowledgments

We thank Jabir Ali Ouassou for useful discussions. This work was supported by the Research Council of Norway through its
Centres of Excellence funding scheme grant 262633 “QuSpin”. N.B. was supported by EPSRC New Investigator Award EP/S016430/1.

\end{document}